\documentstyle[11pt,paspconf]{article}
\def\spose#1{\hbox to 0pt{#1\hss}}
\def\lta{\mathrel{\spose{\lower 3pt\hbox{$\sim$}}
    \raise 2.0pt\hbox{$<$}}}
\def\gta{\mathrel{\spose{\lower 3pt\hbox{$\sim$}}
    \raise 2.0pt\hbox{$>$}}}
\begin{document}

\title{Conference Summary}
\author{Tim de Zeeuw}
\affil{Leiden Observatory}

\bigskip
\bigskip
%\begin{abstract}
%
%\end{abstract}

\keywords{summary}

\noindent
The participants at this conference can broadly be divided into three
separate groups, working on the {\sl stellar} halo, the {\sl gaseous}
halo, or the {\sl dark} halo of our Galaxy. The organizers did an
excellent job of bringing these groups together, and mixing the
various topics throughout the sessions in such a way as to generate
lively discussions. Together with the high-quality posters, this
ensured there was much to learn for everyone. Summarizing all the
material is a daunting task, but when Jeremy Mould pointed out what
appeared to be a freshly-dug grave on Mount Stromlo the other day, and
Phil (Tuco) Maloney reminded me of some memorable scenes in `The Good,
the Bad, and the Ugly' a few hours later, it was clear that there was
no way out. I will discuss the luminous halo first, then review what
we learned about the dark halo, touch briefly on the larger picture,
and close with a look towards the future, with emphasis on the role of
space astrometry missions.

\section{Luminous Halo}

The stellar halo of the Galaxy contains only a small fraction of its
total luminous mass, but the kinematics and abundances of halo stars,
globular clusters, and the dwarf satellites contain imprints of the
formation of the entire Milky Way (e.g., Eggen, Lynden--Bell \&
Sandage 1962; Searle \& Zinn 1978).

\subsection{Field stars: abundances and ages} 

We have heard results from two long-term systematic programs on halo
stars. The first is by Carney, Latham \& Laird, and is based on 1450
kinematically selected stars for which radial velocities, proper
motions, and abundance information have been collected, and orbits
have been calculated. Beers, Norris and Ryan reported preliminary
results based on a sample of 4660 objects from the HK Survey (Beers,
Preston \& Shectman 1992), for which radial velocities, broad-band
colors and abundance indicators are available, but no proper motions
(yet). These studies show that very few really metal--poor stars
([Fe/H] $<-2.0$) occur in the halo. Below [Fe/H]$=-2.0$ the numbers
decrease by about a factor of 10 for every dex in [Fe/H]. Norris
showed that despite much effort, and some false claims, there seems to
be no evidence for stars significantly more metal poor than
[Fe/H]$=-4.0$.

Measurements of detailed element ratios for [Fe/H]$\lta -3.0$ reveal a
significant range in C and N abundances at fixed [Fe/H], which may
well reflect the localised shotnoise of individual enrichment events
caused by early supernovae (e.g., Audouze \& Silk 1995). This
abundance variation is most evident in $r$-process elements, which
suggests that it was put in place by Type II supernovae during the
earliest epoch of star formation, before there was time for
intermediate mass stars to return $s$-process elements to the
interstellar medium. Laird showed that [O/Fe] measurements support
this. This then suggests that the most metal-poor stars were indeed
all formed in a short time interval (less than 1 Gyr), some 12--14 Gyr
ago.  However, Fujimoto argued that the observed abundance variations
may also be caused by internal pollution during the evolution of very
metal-poor stars.

Independent and more direct measurements of ages come from two
sources.  The HIPPARCOS parallaxes of nearby low-metallicity field
halo stars allow them to be placed very accurately in the
Hertzsprung--Russell diagram. Comparison with theoretical isochrones
then results in ages between 11--13 Gyr (Reid 1997). The direct
measurement of the Th/Eu ratio, which is a radioactive clock, gives
similar ages (Cowan et al.\ 1997), albeit with a somewhat larger
uncertainty.

\subsection{Globular Clusters}

Work continues on the ages and metallicities of globular clusters.
Vandenberg described the latest improvements on the theoretical
stellar models. Sarajedini reminded us that the [Fe/H]-scale has
recently been recalibrated by Carretta \& Gratton (1997).
Fortunately, the correction relative to the earlier scale of Zinn \&
West (1984) is monotonic, but it is non-linear, and does affect not
only the inferred ages of individual globular clusters, but also the
commonly accepted properties of the ensemble of clusters. For example,
the well-known bimodal [Fe/H] distribution of the disk and halo
globular clusters may be much less pronounced than was generally
assumed.

Piotto described results derived from systematic programs of
groundbased and Hubble Space Telescope (HST) observations aimed at
obtaining accurate and homogeneous color-magnitude diagrams that reach
down deep enough to include the key regions that are used for age
determinations.  The main results are that (i) nearly all globular
clusters show a remarkably small age-spread, of about 1 Gyr or less,
(ii) the ages of the halo clusters do not correlate with [Fe/H], and
(iii) there is a hint that the clusters at the largest galactocentric
radii might be slightly younger. The globular cluster ages agree with
those of the oldest and most metal-poor field halo stars, and with
those of the oldest populations seen in the dwarf spheroidal
companions. The debate on the precise value of the age of all these
oldest populations continues (e.g., Mould 1998), but for the purpose
of this summary, I will take it to lie in the range of 12--14 Gyr.

Assuming that the evolutionary tracks will continue to improve, there
are two areas where progress can be made in the next few years. It
became evident during the discussions that the various groups that are
interpreting color-magnitude diagrams need to agree on a methodology,
so that e.g., the same quantities are being measured. Secondly, the
beautiful study of NGC 6397 by King et al.\ (1998) shows that HST
proper motions based on {\tt WFPC2} images taken less than three years
apart allow a clean separation of members and field halo stars, and
provide a much improved color-magnitude diagram, extending
significantly deeper than similar ground-based work (e.g., Cudworth
1997). This approach can easily be extended to other nearby clusters,
in particular to Piotto's large sample for which first-epoch exposures
are already in the HST archive.

\subsection{Ghostly Streams}

Scenarios for galaxy formation suggest that the Galactic halo should
contain substructure: extra-tidal material around globular clusters,
tidally disrupted small satellites, and effects of the interaction of
the Milky Way with the Large and Small Magellanic Clouds.  Persistent
hints for velocity clumping in high-$|z|$ samples of field halo giants
can be found in many papers over the past decade (e.g., Freeman 1987;
Majewski, these proceedings). The discovery of the Sagittarius
dwarf (Ibata et al.\ 1994), provided direct and dramatic evidence for
a fairly massive dwarf galaxy that is in the process of tidal
disruption in the Galactic halo. Mateo showed that the extension of
Sgr can now be traced over 34 degrees and counting. Good kinematics
and distances are needed to analyze this further. Grillmair et al.\
(1995) presented evidence for tidal streamers associated with globular
clusters.  Much work in this general area has been done by Majewski,
who reviewed the evidence for retrograde/direct motion at high/low
$|z|$, and reported that the stars associated with the gaseous
Magellanic Stream (\S 1.4) have now also been found. 

The HIPPARCOS Catalog contains a few hundred local halo stars, and the
accurate absolute proper motions combined with the available radial
velocities allow analysis of their space motions. Chiba showed that
while there are Galactic disk stars with $-1.6<$ [Fe/H] $-1.0$, there
is no sign of the disk for [Fe/H]$< -1.6$, and little evidence for
clumping in velocities (but see Helmi et al.\ 1999). The HIPPARCOS
sample is small, and is somewhat of a mixed bag, so it is not easy to
interpret the data. And, as Moody \& Kalnajs illustrated in a poster
contribution, one has to be very careful with extrapolating local halo
measurements to larger distances.

The hints for substructure have triggered work to develop better
detection methods. The Great Circle Method of Lynden--Bell \&
Lynden-Bell (1995), developed for spherical geometry, suggests that
the satellite dwarf galaxies may not all be on independent orbits, but
together occupy a small number of orbits, i.e., are parts of `ghostly
streams'. Majewski showed that the orbits of the globular clusters
display similar signatures. A natural next step is to apply this kind
of analysis to individual stars with good distances and motions, such
as the Carney, Latham \& Laird (1996) sample. 

Tidal stripping in a spherical potential is reasonably well
understood, and was the main topic of the talk by Johnston.  The
spherical approximation applies at large galactocentric radii $r$, and
predicts a coherent structure in ${\bf r}$ {\sl and} ${\bf v}$ space.
Applications include the tidal tails of globular clusters and the
Magellanic Stream. And as Zhao et al.\ showed in a poster
contribution, if one knows which stars belong to the stream through
independent means, the observed kinematics can be used to constrain
the Galactic potential, but only if proper motions of sufficient
accuracy are available. 

The first results of a systematic observational program for finding
substructure in the inner, flattened, halo, was presented by Harding
et al., based on a strategy derived from N-body simulations of their
`spaghetti' model (which graced the announcement poster for this
meeting).  Helmi \& White presented a powerful analytic formalism to
analyze tidal disruption in a flattened potential.  In this case a
coherent structure remains in ${\bf v}$--space, but not in
configuration space. Picking out disrupted streams is therefore more
difficult, but is possible with good kinematic data which includes
astrometry. The properties of the progenitor can then be inferred from
the current measurements, and in this way one can hope to reconstruct
the merging history of our own Galaxy.

There was much interest in the details of the encounter of the Large
and Small Magellanic Clouds (LMC/SMC) with the Milky Way (MW).
Weinberg showed preliminary results of a detailed analysis based on
N-body simulations and normal mode calculations. He computed the
effect of the LMC on the MW halo and disk, and finds that the LMC can
induce the observed warp as well as lopsidedness in the Milky Way,
provided the LMC is sufficiently massive. In turn, the MW tidal field
puffs up the LMC. Weinberg finds a total mass of the LMC of about
$2\times 10^{10} M_\odot$, considerably larger than previous
estimates, perhaps suggesting it has its own dark halo. Photometric
evidence for a stellar halo around the LMC was presented by
Olszewski. This is relevant for the interpretation of the observed
microlensing events towards the LMC (\S 2.2).

Zhao described a scenario in which the Sgr dwarf was involved in an
encounter with the LMC/SMC at a particular time in the past, which
dropped it into a lower orbit. The a priori probability of this is
modest, but it does tie together a number of apparently unrelated
events, and provides a way for Sgr to survive long enough to be torn
apart only in its current perigalacticon passage. It is intriguing
that we may obtain an independent handle on these orbital issues by
determination of the star formation history in the satellites.  This
may well have been punctuated by short bursts of star formation,
coinciding with close encounters or merging. Clearly a full model is
needed of the LMC/SMC/Sgr interaction with the MW, including both the
stars and the gas (\S 1.4). The data to constrain this is quite
detailed, and this nearest encounter may tell us much.

\subsection{Gas: neutral and ionized}

Many of the participants who work on properties of the gaseous
Galactic halo attended the High Velocity Cloud (HVC) workshop at Mount
Stromlo, just prior to this meeting. For a detailed summary, see
Wakker, van Woerden \& Gibson (1999). A number of key results were
presented again during this Symposium, and I will mention these
briefly.

For many years the distribution of high-velocity gas seemed chaotic,
at least to the non-initiate.  The publication of the
Leiden--Dwingeloo Survey (Hartmann \& Burton 1997), and the imminent
completion of its southern extension with the radio telescope in Villa
Elisa, Argentina, is a milestone for the study of neutral gas in the
halo, as it provides a uniform dataset, with improved sensitivity and
angular resolution. This progress is evident on comparing Wakker's
summary map of the HVC's with the discovery map of the HI in the
Magellanic Stream (Mathewson et al.\ 1974). The ambitious HIPASS
project at Parkes is adding further sensitivity and resolution.  My
understanding of the results presented here is that a coherent picture
is finally emerging in which (i) some HVC's are connected to the HI in
the Galactic disk, (ii) much of it is connected to the Magellanic
Stream, with the beautiful HIPASS data of Putman et al.\ (1998) now
also showing the leading arm predicted by numerical simulations of the
encounter of the Magellanic Clouds with the Milky Way (e.g., Gardiner
\& Noguchi 1996; \S 1.3), and (iii) steady accretion of material
either from the immediate surroundings of the Galaxy (as suggested by
Oort 1970), or from within the Local Group (as advocated by Blitz at
this meeting). Van Woerden and Wakker showed that absorption line
studies and metallicity measurements are---at long last---starting to
constrain the distances of individual clouds.  Open issues include the
possibility of hydrodynamic effects influencing the velocities
(discussed by Benjamin and Danly), and the nature of the dense clumps
seen in the HIPASS data.

Ionized halo gas remains elusive. Kalberla, Dettmar and Danly showed
that X-rays and H$\alpha$ emission indicate the presence of $10^6$ K
gas to $|z| \sim 4$ kpc away from the disk.  Maloney showed that the
constraints on the total amount of this material remain rather weak,
an issue to which I shall return in \S 2.4. The dispersion measures of
pulsars in the LMC may help here, as discussed by Bailes, but the
sample of objects is still small.

\section{Dark Halo}

The main topics discussed were the total mass and extent of the dark
halo, the constraints on the mass of halo objects provided by
microlensing experiments, and the nature of the dark matter.

\subsection{Mass and Extent}

Zaritsky summarized what we know about the mass of the Galaxy. The
rotation curve is well established inside a galactocentric radius of
20 kpc, and constrains the mass profile fairly accurately.  Outside
this radius the main constraints are distances and radial velocities
of the globular clusters, the dwarf satellites, and M31. A variety of
methods and arguments show that all measurements to date are
consistent with a model in which the mass distribution is essentially
an isothermal sphere with a constant circular velocity $v_c$$\sim$ 180
km/s. Out to a distance of $\sim$300 kpc---nearly halfway to
M31---this corresponds to a mass of about $2\times 10^{12}
M_\odot$. The average mass-to-light ratio is over 100 in solar units,
so most of this matter is dark, or at least severely underluminous 
(\S 2.4).

The uncertainties in the halo mass profile remain significant, not
only at the largest radii, but even inside the orbit of the LMC. This
affects the determination of the mass fraction of MACHOs in the halo
(\S 2.2). Substantial improvement will have to await better distances
and in particular more accurate absolute proper motions for the
distant globular clusters and satellites. This is not easy, as the
required accuracy for Leo I and II is about 10 $\mu$as/yr (but see
\S~4). The derived space motions would leave only the nature of the
orbits as uncertainty in the determination of the Galactic
potential. The fact that some satellites and clusters may actually be
on the same orbit is a complication.

\subsection{Microlensing}

Enormous effort world-wide is being put into microlensing studies of the
Galactic Bulge, the LMC, and the SMC. Alcock summarized this in a
public evening lecture. At the Symposium, he, Stubbs, and Perdereau
gave more detailed status reports of the MACHO and EROS projects.  The
main triumph of these projects, derived from the 20 events seen to
date towards the LMC, is that point-like objects in the mass range
$10^{-7} \lta M/M_\odot \lta 10^{-2}$ can at most form a minor
constituent of the halo, and hence do not form the bulk of the dark
matter that is tied to the Galaxy. This eliminates most objects of
substellar mass, including planets as small as the Earth.
Furthermore, the events seen towards the LMC have a fairly narrow
duration distribution, which differs from that seen towards the
Galactic Bulge. If these events are caused by halo objects, they must
have masses in a rather narrow range around 0.5 $M_\odot$. Bennett
showed how microlensing events which deviate from the standard
lightcurve may be used to constrain the nature of the lenses further,
and he reminded us that a major uncertainty is the unknown binary
fraction in the lens population (e.g., DiStefano 1999).

The precise MACHO mass fraction of the halo is not easy to determine,
not only because the number of observed events is still modest, but
also because the halo mass distribution out to the distance of the LMC
is not known very well (\S 2.1). It is possible that the MACHOs are
not in the dark halo at all. Flaring of the disk and/or the warp of
the Milky Way, or the presence of another intervening object have been
considered, but it now seems unlikely that these can provide the
entire set of observed events (e.g., Gyuk, Flynn \& Evans
1999). Weinberg's theoretical models of the LMC/MW interaction, and
Olszewski's analysis of the color-magnitude diagram of the LMC
indicate that self-lensing by the LMC, in particular by its own
stellar halo, may well be quite significant.  Their results allow the
possibility that all the lenses are in the LMC itself, as suggested
already by Sahu (1994). This would require lens masses smaller than
0.5 $M_\odot$, which is plausible.  However, if this is the
explanation for the observed events, it then remains a puzzle why the
duration distribution is different from that in the Bulge. The plans
for the future outlined by Stubbs in the preceding talk address these
issues in more detail.

A `byproduct' of the microlensing surveys are massive homogeneous
samples of variable stars in the LMC/SMC and also in the Bulge.  These
are a veritable gold mine for constraining stellar models, and for
tracing Galactic structure. For example, Minniti showed that the RR
Lyrae samples in the Bulge allow an accurate determination of the
luminosity profile of the inner halo to very small galactocentric
radii.

\subsection{How many kinds of dark matter?}

Turner presented his beautifully illustrated `Audit of the Universe'.
This has seen considerable improvement in the past year driven by new
observations. In particular, the distant supernovae projects (Schmidt
et al.\ 1998; Perlmutter et al.\ 1999) indicate a non-zero
cosmological constant $\Lambda$. Turner writes the total mass and
energy density as $\Omega_0=\Omega_M + \Omega_\Lambda$, and finds (to
within factors of less than two, and in units of the critical
density):

\vskip -0.6truecm

\begin{table}
\begin{center}
\begin{tabular}{lll}
\noalign{\smallskip}
$\Omega_{\rm lum}$    &$\sim 0.003$    &(observed) \\
$\Omega_{\rm baryon}$ &$\sim 0.03$     &(Big Bang nucleosynthesis) \\
$\Omega_M$            &$\sim 0.3$      &(galaxy clusters) \\
$\Omega_{\Lambda}$    &$\sim 0.66$     &(supernovae) \\
$\Omega_0$            &$\sim 1.0$      &(anisotropy of cosmic 
                                          background radiation)\\
\end{tabular}
\end{center}
\end{table}

\vskip -0.6truecm

\noindent
This means that while the baryon density $\Omega_{\rm baryon}$ is ten
times larger than the density of luminous matter $\Omega_{\rm lum}$,
it is still only ten percent of the total matter density of the
Universe $\Omega_M$. The `dark energy' $\Omega_\Lambda$ brings
$\Omega_0$ to the critical value within the uncertainties. As Turner
pointed out, this state of affairs raises a number of fascinating
questions, two of which are most relevant for this conference.  If
$\Omega_{\rm baryon}$ is in fact in galaxy halos, which is plausible
given that $\Omega_{\rm galaxies}$ is of the same order, then
non-baryonic dark matter is needed at the scales of clusters and
larger in order to make up the difference between $\Omega_{\rm
baryon}$ and $\Omega_M$. What is this material?  If, on the other
hand, we want only one kind of dark matter on all scales (e.g., for
reasons of simplicity), then the question becomes: where are all the
baryons?  These questions lead directly to the issue of dark matter
composition, to which I now turn.

\subsection{Composition}

Many talks were devoted to the nature of the dark matter. With a few
notable exceptions, most speakers were more sure about what the dark
mass is not made of than about what it is, a conclusion also reached
by Silk, who managed to summarize most of these contributions on the
first day of the conference, i.e., before they were presented!

\smallskip
\noindent
{\sl Brown dwarfs}. Tinney and Flynn reported on programs aimed at
establishing the number density of low-mass objects in the halo, using
groundbased and HST observations. The halo samples contain objects at
a typical distance of 2 kpc, with a tail extending beyond 10 kpc.  The
observed mass functions extend somewhat below 0.08 $M_\odot$, which
marks the transition from stars to substellar objects. The number
density increases with decreasing mass, and perhaps turns over at the
lowest masses---in agreement with the microlensing results (\S
2.2). In any case, smooth extrapolation of the measurements to lower
masses shows that these objects cannot provide all the dark mass in
the Galactic halo. A caveat is that the results are based on
calibrations of local disk objects, which presumably are more
metal-rich than field halo objects.

\smallskip
\noindent
{\sl Compact objects}. Much work was done in the past two years to
investigate whether white dwarfs could be the major constituent of the
dark halo. This activity was triggered by the microlensing events seen
towards the LMC, which suggest a typical lens mass around 0.5
$M_\odot$, and the white dwarf nature of dark matter was defended with
great vigor by Chabrier. However, it seems that in order to have white
dwarfs in sufficiently large numbers requires a special initial mass
function early-on, i.e., a non-standard star formation history. This
cannot be ruled out a priori, but the resulting inevitable
metal-enrichment is hard to hide (Gibson \& Mould 1997). Flynn showed
that HST has not seen these white dwarfs, but perhaps one needs to go
even fainter. Goldman reported that a proper motion survey being
carried out by the EROS team will settle this issue by providing a
strong local constraint. Silk briefly discussed the possibility that
the dark halo consists mostly of neutron stars, and concluded that
these suffer from similar problems as the white dwarfs, and
furthermore, their masses are inconsistent with the microlensing
results.

\smallskip
\noindent
{\sl Cold gas}. In the past few years, a number of authors have
suggested that perhaps the dark matter consists of ultracold (4K)
clumps of H$_2$, with masses of about $10^{-3} M_\odot$, diameters of
30 AU, and densities of $10^{10}$ cm$^{-3}$ (e.g., Pfenniger et al.\
1994; Gerhard \& Silk 1996).  The current microlensing experiments are
not sensitive to such clumps, as they are extended objects rather than
effective point masses. They have never been observed directly in the
local interstellar medium. There is strong disagreement on the
presumed location of this material.  Pfenniger suggested that this
dark matter is in a large outer disk, while Walker proposed that it is
in a spheroidal halo. This latter suggestion has the advantage that
the ionized and evaporating outer envelopes of these clumps could be
responsible for the so-called extreme scattering events seen in radio
observations (Fiedler et al.\ 1987). While Chary reported that the
expected cosmic-ray induced gamma rays are not seen, recent work by
Dixon et al.\ (1998) suggests that they are evident in the EGRET data.

\smallskip
\noindent
{\sl Ionized Gas}. Kahn \& Woltjer (1959) established the mass of the
Milky Way and M31 through their famous timing argument, and suggested
that most of the unseen mass could be ionized gas of about $10^6$ K,
which is very hard to detect.  Maloney summarized the best
observational constraints on the total amount of such gas, and showed
that the limits are no stronger than they were 40 years ago. Kalberla
showed recent ROSAT evidence for such gas, and derived a scale-height
of about 4 kpc. The very sensitive H$\alpha$ surveys that are now
possible with TAURUS-2 (Bland--Hawthorn et al.\ 1998) and with the
WHAM camera (Tufte et al.\ 1998) should allow a measurement of the
total amount of ionized gas in the near future. This may be the best
bet for baryonic dark matter attached to the Galaxy (cf.\ Fukugita, 
Hogan \& Peebles 1998).

\smallskip
\noindent
{\sl Nonbaryonic dark matter}. All of the above candidates for the
dark matter are baryonic. Assuming Turner's audit is correct, this can
make up only 10\% of the total amount of matter in the Universe. It is
natural to assume that this material is associated with galaxies.  The
remaining 90\% of the mass in the Universe (the difference between
$\Omega_{\rm baryon}$ and $\Omega_M$) then has to be made up of
non-baryonic material. Sadoulet, Silk and Turner discussed the
candidates, which include massive neutrinos, axions and neutralinos.
It seems unlikely that the neutrinos provide all the unseen mass on
large scales, because (i) they would not constitute the cold dark
matter that is currently favored by theories of structure formation,
and (ii) the required neutrino mass of $\sim$25 eV seems to be ruled
out by experiments.  To date, there is no experimental evidence for
axions or neutralinos, but the laboratory sensitivity is expected to
improve considerably in the coming year. And finally, Silk argued that
perhaps primordial black holes are the culprit, a most fascinating
suggestion.

\section{Other Galaxies} 

Surface photometry of edge-on disk galaxies to $V$$\sim$28
mag/arcsec$^2$ by Morrison and by Yock shows that these systems
display a variety of luminosity profiles perpendicular to the
disk. This may indicate an outer bulge, a thick disk, or perhaps a
luminous stellar halo. It will be very interesting to try to go to
fainter limiting magnitudes, and to enlarge the sample to search for
correlations with e.g., Hubble type. Determination of the mass
distribution in other disk galaxies is based mostly on HI rotation
curves, and gives strong evidence for extended halos of dark matter,
consistent with the findings for our own Galaxy. Unfortunately,
Kalnajs fell ill, and was unable to present his views on this
topic. Bland--Hawthorn showed convincing evidence that H$\alpha$
measurements can now probe the mass distribution beyond the HI edge,
even though the effort required can only be described as heroic.

The luminous halos of galaxies in the Local Group can be studied in
more detail. Sarajedini showed color-magnitude diagrams for ten
globular clusters in M33, obtained with HST. While some of these
clusters have the same age as the Galactic globular clusters, and are
just as metal-poor, others are several Gyr younger, and have
intermediate metallicities, indicating a more extended halo formation
process. Freeman showed that M31 is very similar to the Milky Way,
albeit a little more massive and with a larger bulge. The kinematics,
abundances, and ages of the M31 globular clusters are very similar to
those in the Milky Way, suggesting they must also have formed quickly.
However, the field stars in the M31 halo are decidedly more metal-rich
than those in the Galaxy. It is not clear how this comes about,
especially since a significant fraction of the halo field stars must
have been tidally dislodged from clusters. The kinematics of the M31
field halo stars should contain further clues to their formation, and
can now be studied in detail through the set of over 1000 planetary
nebulae for which radial velocities have been measured.  Still closer
in, Da Costa and Olszewski showed that satellite dwarf spheroidals and
the Magellanic Clouds all have experienced different star formation
histories, as is evident from the beautiful variety of color-magnitude
diagrams. With the possible exception of the SMC, the oldest
populations invariably have the same age as the Galactic globular
clusters. Clearly, around 12--14 Gyr ago the first generation of stars
was formed synchronously throughout the entire Local Group.

\section{Observational prospects: towards a stereoscopic census}

The currently popular formation scenario was summarized by Wyse and by
White. It assumes a hierarchical build-up of structure in a cold dark
matter universe, with the baryons collecting inside dark halos, and
forming disks. Elliptical galaxies and bulges then result from major
mergers, while disk galaxies remain disks only as long as they
steadily accrete no more than small satellites (e.g., Baugh, Cole \&
Frenk 1996). This scenario then suggests there should be signs of
ongoing accretion and fossil substructure in the Galactic halo, as is
indeed observed. The next step is to test this formation scenario
quantitatively through a detailed comparison with the observed
properties of the Galaxy, and notably its halo. This is a crucial
complement to high-redshift studies of galaxy formation, and should
provide the detailed formation history of our Galaxy.

The theoretical tools to analyse halo substructure that are now being
developed (e.g., Helmi \& White 1999) demonstrate the need for
accurate kinematic data for large samples of halo objects. The ongoing
Hamburg/ESO objective prism survey, outlined in a poster by
Christlieb, covers 10000 square degrees and promises to extend the HK
Survey (\S 1.1) to about 20000 candidate halo objects. Multi-object
spectroscopy by 2DF or SLOAN will provide radial velocities and
abundances (see poster by Pier). The HIPPARCOS Catalog contains
globally accurate proper motions with accuracies of $\sim$1 mas/yr for
a few hundred bright and nearby halo stars.  Combination of the Tycho
positions for nearly 3 million stars to $V$$\sim$12 with those in the
Astrographic Catalog, which are each of modest individual accuracy but
have an epoch difference of about 80 years, will provide proper
motions of 2--3 mas/yr (Hoeg et al.\ 1998). The resulting TRC/ACT
database will contain about 30000 halo stars, and should be available
in a year or two. As 2 mas/yr translates to 10 km/s at 1 kpc, this
will allow space motions to be derived for halo stars out to distances
of a few kpc. The USNO-B Catalog reaches about five magnitudes
fainter, and can be put on the HIPPARCOS/Tycho reference system, but
the accuracies of individual proper motions will be only $\sim8$
mas/yr (Monet 1997), making them of modest value for the study of the
Galactic halo.
 
The success of HIPPARCOS has resulted in further interest in space
astrometry, leading to the ongoing ESA study of a mission-concept
called GAIA (e.g., Gilmore et al.\ 1998). Interest in imaging
terrestrial extrasolar planets is pushing NASA's Space Interferometry
Mission. SIM will be a pointed observatory, and will provide proper
motions and parallaxes of micro-arcsecond ($\mu$as) accuracy by the
end of the next decade. If approved, GAIA will provide parallaxes and
absolute proper motions of similar quality for all the one billion
stars to $V$$\sim$20, together with radial velocities and photometry
or low-resolution spectroscopy for most objects brighter than
$V$$\sim$17, by 2015. This is a long time, but the tremendous gain
over HIPPARCOS, summarized below, will be worth the wait.

\bigskip
\centerline{Comparison of HIPPARCOS and GAIA}
\vskip -5.0truemm
\begin{table}
\begin{center}
\begin{tabular}{llll}
\tableline\tableline
\noalign{\smallskip}
                  & HIPPARCOS        & GAIA \\
\tableline
\noalign{\smallskip}
Magnitude limit   &  $\sim 12$       &$V=21$ \\
Completeness      &7.3--9.0          &$V=20$ \\
Number of objects &$1.2\times 10^5$  &$1.2\times 10^9$ \\
Accuracy          &1--2 mas          & 4 $\mu$as  &($V<10$) \\
                  &                  &10 $\mu$as  &($V=15$) \\
                  &                  &0.2 mas     &($V=20$) \\
Radial velocities &--                &3--10 km/s  &($V$$<$16--17)\\
Narrow-band photometry &--           &multi-color &($V$$<$16--17)\\
or low-resolution spectroscopy &--   &$R$=20--40\AA &($V$$<$17--18)\\
\noalign{\smallskip}
\tableline
\end{tabular}
\end{center}
\end{table}

\vskip -4.0truemm
\noindent
This is not the place to summarize the entire scientific rationale for
$\mu$as astrometry from space, and its impact on all aspects of the
structure of the Galaxy, stellar physics, detection of planets around
other stars, etc. The relevance for the study of the Galactic halo is
best illustrated by considering the mind-boggling accuracy of the SIM
and GAIA proper motions: an uncertainty of 10$\mu$as/yr translates to
5 km/s at 100 kpc! This will make it possible to measure the space
motions of globular clusters and all the dwarf satellites, including
Leo I and II. The unbiased coverage of the entire sky by GAIA will
allow identification of very large halo samples from the narrow-band
photometry, and provide measurements of the full six-dimensional phase
space information (positions and velocities) throughout the inner halo
(to distances of about 20 kpc from the Sun), and full velocity
information on individual stars to much larger distances.  This will
e.g., (i) provide the mass distribution of the Galaxy with
unprecedented accuracy, (ii) allow kinematical selection of the
members of the globular clusters and dwarf satellites, leading to
clean color-magnitude diagrams, and (iii) trace ghostly streams and
substructure.  High-resolution spectroscopic follow-up will provide
the distribution of the abundances throughout the halo.  This will
allow reconstruction of the full formation history of the Galaxy.

\section{Concluding remarks}

This Symposium is dedicated to Alex Rodgers.  Alex was well-ahead of
the field with his early work on metal-rich halo stars (Rodgers,
Harding \& Sadler 1981), and his investigation on retrograde globular
cluster orbits (Rodgers \& Paltoglou 1984), both of which indicated
the presence of infall and substructure. He would clearly have enjoyed
this meeting. The string of three succesful and stimulating
international Stromlo Symposia have established the series, and are a
credit to Australian astronomy. We are all looking forward to number
four!

\acknowledgments 
It is a pleasure to thank Jeremy and Joan Mould for warm hospitality
during the workshop, and for introducing me to the Red Belly Black
Cafe. The Leids Kerkhoven Bosscha fund kindly provided a travel grant.
Ken Freeman, Amina Helmi and John Norris commented on an earlier
version of the manuscript.

\end{document}